\input harvmac.tex
\def\tx{\tilde X}

\def\tv{\tilde V}
\def\tz{\vec{\tilde\zeta}}
\def\vz{\vec \zeta}
\def\sa{\Sigma_5^A}
\def\sb{\Sigma_5^B}

\def\tj{\tilde J}
\Title{\vbox{\baselineskip12pt\hbox{hep-th/9710124}
\hbox{RU-97-80}}}
{\vbox{
\centerline{Neveu-Schwarz Five-Branes And}
\vskip 10pt
\centerline{Matrix String Theory On K3}
}}
\centerline{Duiliu-Emanuel Diaconescu and Jaume Gomis}
\medskip
\centerline{\it Department of Physics and Astronomy}
\centerline{\it Rutgers University }
\centerline{\it Piscataway, NJ 08855--0849}
\medskip
\centerline{\tt duiliu, jaume@physics.rutgers.edu}
\medskip
\bigskip
\noindent

The Matrix theory description of Type IIA string theory on a
compact K3 surface as  the theory of Neveu-Schwarz
five-branes  on $\tilde{K3}\times S^1$ is analyzed. The full
multiplet of space-time BPS states is identified  in the
five-brane world-volume as fluxes.

\bigskip

\Date{October 1997}

\newsec{Introduction}
The  Matrix description of M-theory \ref\BFSS{T. Banks, W. Fischler,
S. H. Shenker, L. Susskind, Phys.Rev. {\bf D55} (1997) 5112, 
hep-th/9610043.} on $T^d$ is 
$SYM_{d+1}$ on the dual torus $\tilde{T^d}$
\nref\WT{W. Taylor, Phys.Lett. {\bf B394} (1997) 283,
hep-th/9611042.}%
\nref\GRW{O. J. Ganor, S. Ramgoolan, W. Taylor,
Nucl.Phys. {\bf B492} (1997) 191, hep-th/9611202.}%
\refs{\BFSS, \WT, \GRW}.
This definition
is incomplete for $d>3$ since  then the SYM theory is not renormalizable and
more data needs to be added to describe the full theory.
Compactifications on $T^4$ and $T^5$ have recently been proposed in
terms of  the five-branes of M-theory and Type II string theory
\nref\Roz{M. Rozali,  Phys.Lett. {\bf B400} (1997) 260,
hep-th/9702136.}%
\nref\BRS{M. Berkooz, M. Rozali, N. Seiberg, hep-th/9704089.}%
\nref\S{N. Seiberg, hep-th/9705221.}%
\refs{\Roz, \BRS, \S}.
A unified derivation for matrix theory compactification on tori has
been given in 
\nref\Seib{N. Seiberg, hep-th/9710009.}%
\refs{\Seib}.
The Matrix theory description of curved 
backgrounds has been given in 
\nref\Doug{M. R. Douglas, hep-th/9612126.}%
\nref\FR{W. Fischler, A. Rajaraman, hep-th/9704123.}%
\nref\DG{D.E. Diaconescu, J. Gomis, hep-th/9707019.}%
\refs{\Doug, \FR, \DG}
for non-compact ALE manifolds and in 
\nref\GovA{S. Govindarajan, hep-th/9705113.}%
\nref\BR{M. Berkooz, M. Rozali, hep-th/9705175.}%
\nref\GovB{S. Govindarajan, hep-th/9707164.}%
\refs{\GovA, \BR, \GovB}
for compact K3 surfaces.
The former approach based on the quiver quantum mechanics 
of 
\nref\MD{M.R. Douglas, G. Moore, hep-th/9603167.}%
\nref\JM{C. V. Johnson, R. C. Myers, Phys.Rev. {\bf D55}
(1997) 6382, hep-th/9610140.}%
\refs{\MD, \JM} must be modified for the compact case.
Seiberg has proposed in \refs{\S} that the Matrix
theory description of Type IIA string theory on K3 is given by the
theory of Neveu-Schwarz (NS) five-branes compactified on
\foot{The relation between
these manifolds will be made precise in section 4.} $\tilde{K3}\times
S^1$. In this paper we 
make this proposal more precise by identifying the expected space-time
BPS states in the NS five-brane theories. In section 2 we consider Type IIB NS
five-branes on $\tilde{K3}\times S^1$. We analyze possible bound
states with D-branes and find fluxes in the five-brane theory
corresponding to the bound states.
In section 3 a similar treatment is given for Type IIA NS five-branes. In
section 4 we give a Matrix interpretation to these wrapped brane
theories. We show how T-duality acts on K3  and  identify the $SO(4,20,Z)$
multiplet of BPS states of Type IIA string theory on K3.

\newsec{Type IIB Neveu-Schwarz five-branes on $K3\times S^1$}

In this section we consider the world-volume theory of $N$ IIB NS
five-branes compactified on a K3 surface
\foot{This problem has been considered before in 
\ref\DVVb{R. Dijkgraaf, E. Verlinde, H. Verlinde,  Nucl.Phys. {\bf
B484} (1997) 543, hep-th/9504047.}
for different reasons.} $\tilde{X} \times S^1$. 
S-duality \refs{\S}
shows that at energies below the string scale $M_s$ the 
flat volume theory is $(1,1)$ $SYM_{5+1}$ with gauge group $U(N)$ 
and gauge coupling 
\eqn\gcoupl{
{1\over g^2}=M_s^2.}
When compactified on $\tilde{X}\times S^1$
 one is left with eight real unbroken supercharges. Excitations of
this system can be described by bound states of wrapped D-branes with  NS
five-branes.  In the $g_s \rightarrow 0$ limit, where string modes
decouple, these excitations can be interpreted as solitonic objects of the
$SYM_{5+1}$ on $\tilde{X} \times S^1$ theory. A detailed 
understanding  of these bound states can be obtained by analyzing the
S-dual description \foot{This is justified since we will be
interested in BPS 
states only.} consisting of $N$ D5-branes compactified on
$\tilde{X}  \times S^1$. In 
\nref\MRD{M.R. Douglas, hep-th/9512077.}%
\nref\GHM{M. Green, J.A. Harvey, G. Moore, Class.Quant.Grav. {\bf 14}
(1997) 47, hep-th/9605033.}%
\nref\HM{J.A. Harvey, G. Moore, hep-th/9609017.}%
\refs{\MRD, \GHM, \HM} it was shown that the world-volume
theory has the Chern-Simons coupling
\eqn\cscoup{
I_{CS}=\int_{\tilde{X}\times S^1}C\wedge ch(F)\sqrt{\hat{A}(R)}.}
$C$ is the sum of all RR vector potentials, $ch(F)=\hbox{Tr}e^{{iF
\over 2\pi}}$ is the Chern-character and $\hat{A}(R)$ encodes the
information of the curvature of the brane. From this coupling one can
read off the charges under the different RR vector potentials in
terms of the $\hbox{rank}(F)$, $c_1(F)$ and $c_2(F)$ of a vector
bundle over $\tilde{X}\times S^1$. In \refs{\HM}
the moduli space of D-branes was identified with the
moduli space of vector bundles \foot{Actually string duality requires
these moduli 
spaces to be 
compactified by adding semi-stable coherent sheaves \refs{\HM}. This
distinction  
is unnecessary in the present paper.} on $\tilde{X} \times S^1$ with
specified topological data. 
Therefore, BPS states are
represented by harmonic forms  on the moduli space of vector
bundles on  $\tilde{X} \times S^1$. 
More precisely a BPS configuration with charge vector 
\eqn\chargeA{
Q=\left(r,l,s\right)\in H^0(\tx,Z)\oplus H^2(\tx,Z)\oplus H^4(\tx,Z)}
is described by a $U(r)$ instanton bundle with Mukai vector 
\eqn\mukai{
Q=v\equiv\left(r,c_1,{1\over 2}c_1^2+r-c_2\right).}
The integers $(r,l,s)$ represent the net D5, D3 and D1 charge
respectively. The dimension of the moduli space is positive only for 
states satisfying the BPS condition
\eqn\BPS{
v^2\equiv l^2-2rs\geq -2,}
where the inner product on $H^{2*}$ is given by the intersection form.
Note that a wrapped D5-brane on
$\tilde{X} \times S^1$ induces an effective negative D1-brane
charge \ref\eff{M. Bershadsky, V. Sadov, C. Vafa, Nucl.Phys. {\bf B463}
(1996) 420, hep-th/9511222.}. In the presence of wrapped states, when
$c_1\neq 0$, there is a further contribution to the
induced negative D1-brane charge.

In the S-dual picture one has  NS five-branes instead
of D5-branes  and $r,l,s$ correspond to the net NS five-brane
charge, D3-brane charge and
fundamental string winding number respectively. The energy of a state
with charge vector \chargeA\ is given by the following BPS mass
formula \foot{Generally, the mass formula also depends on the periods of the 
NS two-form field. We will consider $B=0$ throughout the paper.}
\eqn\energA{E=\sqrt{\left({1\over g_s^2}r\tilde V\Sigma_5^BM_s^6
+s\Sigma_5^BM_s^2\right)^2
+\left({1\over g_s}l\Sigma_5^B|\tz |M_s^4\right)^2}.}
Here we  have used the well known BPS mass formulas for wrapped branes and 
fundamental string windings in Type IIB string theory. $\Sigma_5^B$ is
the circumference of the circle and $\tilde{V}$ is the volume of
$\tilde{X}$. The vector
$\tz$ represents the periods of the K\"ahler  
class $\tilde J$ and holomorphic two form $\tilde\Omega$ on the cycle
$\gamma \in \tilde{X}$ which the D3-branes wrap
\eqn\periods{
\tz=\left(\int_\gamma \tilde J,\ \int_\gamma\tilde\Omega\right).}

We will consider first the binding energy of $r=N$ NS five-branes and $n$
D1-branes and D5-branes respectively. Note that the theory has
different topological sectors labeled by the instanton number $c_2$.
It will turn out later that the relevant sectors for matrix theory 
correspond to $c_2\geq N$. Here we consider for simplicity $c_2=N$
so that the net string winding number cancels. 
In the limit in which the brane
physics decouples from the bulk one has the following energies 
\eqn\energD{
E_1=\lim_{g_s \rightarrow 0}\sqrt{\left({1\over g_s^2}N\tilde
V\Sigma_5^B M_s^6\right)^2 
+\left({1\over g_s}n\sb M_s^2\right)^2}-
{1\over g_s^2}N\tv\Sigma_5^BM_s^6=
{n^2\over 2N}{\sb\over \tv M_s^2}}
for $N$ NS five-branes and $n$ D1-branes and 
\eqn\energE{
E_5=\lim_{g_s \rightarrow 0}\sqrt{\left({1\over g_s^2}N\tilde
V\Sigma_5^B M_s^6\right)^2+ 
\left({1\over g_s}n\tv\sb M_s^6\right)^2}-
{1\over g_s^2}N\tv\Sigma_5^BM_s^6=
{n^2\over 2N}\tv\sb M_s^6}
for $N$ NS five-branes and $n$ D5-branes \foot{Here we have to tune
$c_2=n$ for the D5-brane vector bundle  
in order to cancel the induced D1-brane charge.}.

In a similar fashion to \refs{\S} one can identify the $E_1$
excitation with the energy of the electric flux in the $U(1)$ part of the
$U(N)$ curvature of the NS five-branes theory 
\nref\DVVa{R. Dijkgraaf, E. Verlinde, H. Verlinde,  Nucl.Phys. 
{\bf B486} (1997) 89, hep-th/9604055.}%
\refs{\DVVa, \BRS, \S}. 
The energy of $n$ units of electric flux around $S^1$ is 
\eqn\energF{
E_1={n^2\over 2N}{\sb\over \tv M_s^2},}
which matches with the binding energy \energD\ . As already noted in
\refs{\S} the binding energy $E_5$ cannot be identified as an
excitation in the $SYM_{5+1}$.

We consider now the bound state of $N$ NS five-branes and $n$ D3-branes
wrapped on $\gamma \times S^1$. These configurations correspond to
$U(N)$ bundles with $c_1 \neq 0$. The relevant topological sectors
are labeled by the instanton number 
\eqn\instnum{
c_2-{1 \over 2} c_1^2=N+s,\qquad s\geq 0.}
Equivalently, we can label them by the net fundamental string winding 
charge $s$. In order to satisfy the BPS condition \BPS\ for arbitrary 
$n$ and arbitrary cycles we have to consider sectors with high enough $s$. 
The matrix theory interpretation will be clarified in section 4. 
The energy is in this case 
\eqn\energB{\eqalign{
E_3=&\sqrt{\left({1\over g_s^2}N\tilde V\Sigma_5^B
M_s^6+s\Sigma_5^BM_s^2\right)^2+\left({1\over g_s}n|\tz |\Sigma_5^B
M_s^4\right)^2}\cr & -{1\over g_s^2}N\tv\Sigma_5^BM_s^6
-s\sb M_s^2.\cr}}
In the limit $g_s\rightarrow 0$ this becomes 
\eqn\energC{
E_3={n^2\over 2N}{|\tz|^2\Sigma_5^BM_s^2\over \tv}.}

In order to understand the flux associated with wrapped three-branes
we return  to the S-dual description. Consider a bound
state of $N$ D5-branes and $n$ D3-branes wrapped on 
$\gamma \times S^1$. As before we add background D1-branes wound
around $S^1$ to cancel the net D1 charge. This configuration is
described by an $U(N)$ field strength $F$ such that the trace $U(1)$
part $F_1$ is an harmonic form and the non-abelian part $F_0$ is
anti self-dual \HM. These conditions define in fact an Einstein-Hermitian 
structure
\nref\Ko{S. Kobayashi, Proc. Japan. Acad. Ser. A. Math. Sci., {\bf 58}
(1982), 158.}%
\nref\L{M. L\"ubke, Math. Ann. {\bf 260} (1982), 133.}%
\refs{\Ko, \L}.
The low energy effective
action for this sector of string theory is
\eqn\actA{
S=S_{bulk}+S_{D5}+S_{D3}+S_{D1}.}
The relevant Chern-Simons couplings with the RR fields can be read off
from \cscoup
\eqn\actB{
{i\over 2\pi}\int_{R\times S^1\times\tx}\hbox{Tr}F\wedge C^{(4)}+
n\int_{R\times S^1\times\gamma}C^{(4)}}
where $C^{(4)}$ is the RR four-form potential with self-dual field
strength $H^{(5)}_+$. This can be rewritten as
\eqn\actC{
{i\over 2\pi }\int_{R\times S^1\times\tx}\hbox{Tr}F\wedge C^{(4)}+
n\int_{R\times S^1\times\tx}C^{(4)}\wedge\omega}
where $\omega$ is the Poincar\'e dual of $\gamma$.
The equation of motion for $C^{(4)}$ is
\eqn\eqmotion{
d*H^{(5)}_+={i\over 2\pi }\hbox{Tr}F\wedge\delta^{(4)}+
n\omega\wedge\delta^{(4)}}
where $\delta^{(4)}$ is a transverse delta four-form. Integrating 
\eqmotion\ over $S^3\times S^1\times \eta$ with $\eta$ an {\it arbitrary}
two cycle in $\tx$ we find
\eqn\inteq{
\int_{\eta}\left({i\over 2\pi }\hbox{Tr}F+n\omega\right)=0,
\qquad \forall \eta\subset \tx.}
Therefore the $U(1)$ flux induced by the three-brane is
\eqn\fluxA{
F_1={{2\pi in\over N}}\omega I_N.}
The effective D1 charge is induced by the Chern-Simons coupling
\eqn\actD{
-{1\over 8\pi^2}\int_{R\times S^1\times\tx}\hbox{Tr}
F^2\wedge C^{(2)}.}
Therefore the induced charge is given by the Chern-Weil formula
\eqn\quantA{
c_2-{1\over 2}c_1^2={1\over 8\pi^2}\hbox{Tr}F^2.}
The non-abelian flux $F_0$ is determined from \quantA
\eqn\quantB{
{1\over 8\pi^2}\int_{\tx}\hbox{Tr}F_0^2=N+s+{n^2\over 2N}\int_{\tx}
\omega^2.}

We conclude that D3 charged states appear in sectors of the theory
characterized in the low energy limit by different instanton numbers. 
Formula \quantB\ shows that the non-abelian instanton number can
become fractional.
This is not an inconsistency since it is the second
Chern character \quantA\ which must be integral. Equivalently,
\quantB\ is the fractional $SU(N)$ 't Hooft flux in $U(N)$ 
gauge theories.

We will now show that the energy of the magnetic flux in the NS five
brane theory coincides with the binding energy \energC. We rotate the complex
structure of $\tx$ so that the cycle $\gamma$ is holomorphic. Then 
$\omega$ is a $(1,1)$ form whose self-dual part is given by the
projection along the K\"ahler class
\eqn\Kproj{
\omega_+={\int_{\tx}\omega\wedge {\tj}\over \int_{\tx}{\tj}
\wedge \tj}\tj .}
Note that for this complex structure
\eqn\vecper{
\tz= \left(\int_{\tx}\omega\wedge \tj,0\right),
\qquad 
\tv={1\over 2}\int {\tj}\wedge {\tj}.}
The energy of the flux is 
\eqn\energG{
E=-{M_s^2\sb \over 8\pi^2}\int_{\tx}\hbox{Tr}\left(F\wedge*F \right)
={M_s^2\sb \over 8\pi^2}\left({4\pi^2n^2\over N}\left(\parallel \omega_+
\parallel^2+ \parallel \omega_- \parallel ^2\right) +\parallel
F_0\parallel^2\right)} 
where $\parallel \cdot \parallel$ denotes $L^2$ norm.
Using \quantB,\ \vecper\ and 
\eqn\Poinca{
\int_{\tilde{X}}\omega \wedge \omega=\parallel \omega_+
\parallel^2-\parallel \omega_- \parallel^2,}
we find the energy of the total $U(N)$ flux
\eqn\energH{
{n^2\over 2N}{|\tz|^2\sb M_s^2\over \tv}+(N+s)\sb M_s^2.}

The first term is in precise agreement with the binding energy 
\energC.\ The second term represents the energy of the fundamental
strings dissolved in the five-branes as instantons. It appears here
naturally since we have computed the energy of the total $U(N)$ flux.

\newsec{Type IIA NS five-branes on $K3 \times S^1$}

Here we consider the six dimensional theory of Type IIA NS five-branes
on $\tilde{X}\times S^1$. This is related by T-duality on the circle
to the Type IIB theories analyzed in the previous section. After
T-duality,  the Mukai vector $(r,l,s)$ corresponds to net NS five-brane
charge, D2-brane charge and fundamental string momentum around the
circle respectively. It is important to note that winding 
is mapped to momentum of the fundamental string. This suggests that
wrapped NS five-branes  on $\tilde{X}\times S^1$ induce a string
momentum charge around the circle. The excitations of this system can
be described in a similar fashion as bound states with D0, D2 and D4
branes in the free string limit. 

These excitations can also be
identified as fluxes in the NS five-branes theory. Along the moduli
space of this theory, there are $r$ tensor multiplets of $(2,0)$
supersymmetry. Each of them includes five scalars and a two-form B,
whose field strength is self-dual. It is important to note that one of
the scalars is compact, and its radius is proportional to the M-theory
circle yielding Type IIA string theory. The compact scalar $\phi$ is
dualized as in \refs{\S} to a four-form $A_4$ with a five-form field
strength $F_5$. Bound states of $N$ NS five-branes with $n$ D0-branes
can be identified with $n$ units of $F_5$ magnetic flux in
$\tilde{X}\times S^1$. The
corresponding energy is
\eqn\energI{
E_0={n^2\over 2N}{1\over \tv\sa  M_s^4}}
which is T-dual to \energD. Bound states of $N$ NS five-branes with
$n$ D4-branes 
can be identified with $n$ units of $F_5$ electric flux along the
circle. Their energy 
\eqn\energJ{
E_4={n^2\over 2N}{\tv M_s^4\over\sa}}
is related by T-duality to \energE.

The flux associated with the wrapped D2-branes states cannot be
completely described in Type IIA variables since the theory of
non-abelian tensor multiplets is not known. It is however clear that
the binding energy can be read off directly from \energC\ after
T-duality on the circle.

\newsec{Matrix Theory on K3}

Seiberg has proposed
\refs{\S} that the Matrix description of Type IIA on K3 is given by
the NS five-brane theory compactified on $\tilde{X}\times S^1$, where
$\tilde{X}$ denotes the T-dual to the space-time K3 surface $X$. 
In order to have
a complete theory on the brane we send $g_s \rightarrow 0$ and
keep the string scale $M_s$  fixed \foot{We want to keep the degrees
of freedom corresponding to the M-theory membranes wrapped around the
M-theory circle.}. In 
\refs{\Seib} it was shown that the DLCQ quantization of M-theory on a
light-like circle of radius $R$ is the Planck scaled, boosted,
$R_s\rightarrow 0$ limit of another M-theory compactified on a space-like
circle of radius $R_s$. The definition of compactified M-theory
then follows from T-duality on D0-branes
\nref\sen{A. Sen, hep-th/9709220.}%
\refs{\sen, \Seib}. This logic can be applied to
the present situation using a modified version of the T-duality 
transformation for K3 surfaces defined
in \ref\HO{K. Hori, Y. Oz, hep-th/9702173.}. Since this maps a D0-brane
charge on $X$ to a D0-D4 bound state on $\tx$, we obtain naturally a 
sector of the theory characterized by instanton number $c_2=N$ as
anticipated in section 2
\foot{The precise map between BPS states on $X$ and BPS states on
$\tx$ can be defined as Fourier-Mukai transform \HO. To keep the
discussion simple, we do not present the details.}.
The sectors with higher instanton number
contain longitudinal matrix theory five-branes, \GRW,\ which can be
identified with longitudinal heterotic solitons 
\nref\HS{J.A. Harvey, A. Strominger, Nucl.Phys. {\bf B449} (1995) 535,
hep-th/9504047.}%
\nref\PKT{P. K. Townsend, Phys.Lett. {\bf B354} (1995) 247, 
hep-th/9504095.}%
\refs{\HS, \PKT}.
The space-time and auxiliary theory parameters can be related as follows. 
For Type IIA NS five-branes \refs{\S}
\eqn\param{\eqalign{
\Sigma^A_5&={l_p^6 \over RV} \cr
M_s^2&={R^2VL_5 \over l_p^9} \cr}}
and
\eqn\parama{\eqalign{
\Sigma^B_5&={l_p^3 \over RL_5} \cr
M_s^2&={R^2VL_5 \over l_p^9} \cr}}
for Type IIB five-branes. These relations follow naturally from the
analysis in \Seib. $V$ is the volume of the
space-time K3, $L_5$ is the circumference of the circle which gives
the Type IIA dilaton 
from M-theory  and $\Sigma_5$ is the circumference of the circle
which the NS five-branes wrap. 

To identify the fluxes found previously
with space-time states in the infinite momentum frame (IMF) one has to
relate the geometry of $X$ to that of $\tilde{X}$. T-duality on K3
surfaces can be defined as follows \HO.
Recall that the K3 
cohomology lattice can be split as 
\eqn\splitA{
\Gamma^{4,20}\simeq\Gamma^{3,19}\oplus \Gamma^{1,1}.}
The first factor is identified with $H^2(X,Z)$ and the second 
factor, generated by the null vectors $w,\ w^*$, is identified
with $H^0(X,Z)\oplus H^4(X,Z)$
\ref\A{P. S. Aspinwall, In: {\it Mirror Symmetry II}, B. Greene and
S.-T. Yau, Eds., International Press, Cambridge, 1997, hep-th/9404151; 
TASI96(K3), hep-th/9611137.}. 

The
space-like four plane which defines the conformal field theory
decomposes similarly
\eqn\splitB{
\Pi\simeq B^\prime\oplus\Sigma^\prime}
where
\eqn\Bfield{
B^\prime=V w+w^*+B}
determines the volume and the B-field in a standard way \A,
and $\Sigma^\prime\cap{w^{\perp}/w}$ is identified with the 
self-dual three plane 
in $H^2(X,R)$. Here the field $B$ is set to zero.
The splitting \splitA\ follows naturally from compactifying the
M-theory heterotic duality on a circle of radius $R$. This gives rise
to the following relations
\eqn\rela{
\alpha_{A}^{\prime}g_A^2=R^2, \qquad \alpha_{A}^{\prime}={l_p^3 \over R},
\qquad {1 \over  \alpha_{H}^{\prime}}={V \over g_A^2 \alpha_{A}^{\prime 3}}.}
In the heterotic theory, T-duality on the extra circle can be viewed
as an automorphism of the Narain lattice interchanging $w\leftrightarrow
w^*$. Using \rela\ and the usual T-duality formula
\eqn\hetdual{
\tilde R={{\alpha}_H^{\prime}\over R},}
it follows that 
\eqn\IIAdual{
\tilde V={{\alpha}_A^{\prime 4}\over V},\qquad 
{\tilde g}_A^2=g_A^2{{\alpha}_A^{\prime 4}\over V^2}.}
Therefore the K3 volume and the IIA dilaton transform according to 
standard T-duality formulae. We conclude that T-duality on K3 can be
defined as an automorphism of the cohomology lattice exchanging 
$w\leftrightarrow w^*$. This does not affect the second 
cohomology lattice $\Gamma^{3,19}$, hence the dual surfaces have the
same complex structure. 
Note that this transformation is not uniquely defined by \IIAdual.
In fact, 
it can be combined with any automorphism of $\Gamma^{3,19}$
to yield a valid symmetry transformation. In Matrix theory this
freedom represents the redundancy of the description of the base 
theory. 

The K\"ahler class is simply rescaled as 
\eqn\Tduality{\eqalign{
& {\tilde J}={l_p^6\over R^2V}J.\cr}}
Given a two-cycle $\gamma\subset X$ the vectors of periods satisfy the
dimensionless relation
\eqn\rel{
{\tz^2\over \tv}={{\vz}^{2}\over V}.}

Following \refs{\S} we identify the bound state energies
computed previously with BPS states of Type IIA string theory on K3
in the IMF. Since $E_i={RM_i^2\over 2N}$ for particle-like states in the
IMF one has

\noindent
$\bullet$ Type IIB Bound states
\eqn\IMFB{\eqalign{
M_1^2&=\left({1 \over L_5}\right)^2 \cr
M_5^2&=\left({L_5V \over l_p^6}\right)^2 \cr
M_3^2&=\left({|\vec{\zeta}| \over l_p^3}\right)^2\cr}}
$\bullet$ Type IIA Bound state
\eqn\IMFA{\eqalign{
M_0^2&=\left({1 \over L_5}\right)^2 \cr
M_4^2&=\left({L_5V \over l_p^6}\right)^2 \cr
M_2^2&=\left({|\vec{\zeta}| \over l_p^3}\right)^2\cr}}
Let us see what is the space-time interpretation of these states. $M_1,
M_0$ correspond to KK momenta along the $L_5$ circle in the M-theory
language and to D0-branes in the Type IIA language. $M_5, M_4$
correspond to a wrapped M-theory five-brane on $K3\times S^1$ or to a
D4-brane wrapped on $K3$. Finally, $M_3, M_2$ correspond to an
M-theory membrane or a D2-brane wrapped on a cycle $\gamma$ in
$K3$. Since $H_2(X,Z)\simeq \Gamma^{3,19}$ we are able to identify the
expected 
$SO(4,20,Z)$ multiplet of BPS states.

\newsec{Discussion}

In this paper we have identified  fluxes in the NS five-brane
theories corresponding to various bound states of D-branes with NS
five-branes in the free string limit $g_s=0$. Furthermore, using
T-duality on K3, we have been able to give a Matrix theory description
of Type IIA string theory on K3. It is worth mentioning that this
identification makes sense only in the $g_s=0$ limit in which the
physics on the NS five-branes of Type IIA and Type IIB decouple from
the bulk modes.  Even though the
bulk modes decouple, they  give classical backgrounds to the
five-brane theories. The moduli space of compactification data is
therefore $SO(4,20,Z)\backslash SO(4,20)/(SO(4)\times SO(20))$.
This makes manifest the moduli space of
vacua of the space-time theory. In a very similar manner one can give
a Matrix description of M-theory on K3 as the theory of the M-theory
five-brane on $\tilde{K3}\times S^1$. The analysis follows in an
straightforward manner from the one we considered here.

We noted in section 2 that in order to describe wrapped states  the
instanton number of the vector bundle must be shifted. This is
consistent with Douglas's proposal \refs{\Doug}. There it was shown 
that in order to describe wrapped states with nonzero charge 
one has to consider non-regular quiver representations.
As noted in \refs{\DG} this is not necessary
if the total D2-brane charge is zero, so that the original quiver can
describe multiple wrapped brane states whose total charge is zero. It
is nice to note how this conclusion can be reached directly from the
NS five-brane theories. 

These NS five-brane theories are not ordinary local quantum field
theories since the base space is ambiguous. It would be interesting to
identify states of the S-dual space-time theory, providing a Matrix 
definition of heterotic string on $T^4$. Partial evidence for this
comes from the analysis of the zero modes of the
NS five-brane wrapped on K3
\refs{\HS,\PKT} which yields the precise chiral structure of the
heterotic string.

An important step towards proving that the Matrix theory description
of M-theory compactifications is valid requires to check the
degeneracy of BPS states in Matrix theory\foot{We thank M. Douglas for
pointing this to us.}. Partial results along these lines were undertaken
in  \ref\Gopa{R. Gopakumar, hep-th/9704030.} but many more checks need
to be made. For the case we consider, the degeneracy of the fluxes of
a single Type IIA NS five-brane theory was shown in
\DVVb\ to exactly reproduce the degeneracy of BPS
states of Type IIA on K3. It would be nice to generalize this result
to arbitrary $N$. 
\bigskip
\centerline{\bf Acknowledgments}

We would like to thank T. Banks, M. Berkooz, M. Douglas,
L. Nicolaescu, N. Seiberg,
S. Shenker and E. Witten for helpful discussions.

\listrefs
\end